\documentclass[prd, aps, superscriptaddress, twocolumn, preprintnumbers, floatfix, nofootinbib]{revtex4}

\pdfoutput=1    
\usepackage{epsf}
\usepackage{amsmath}
\usepackage{amssymb}

\usepackage{graphicx}
\usepackage{dcolumn}
\usepackage{bm}
\def\be{\begin{equation}}
\def\ee{\end{equation}}
\def\bea{\begin{eqnarray}}
\def\eea{\end{eqnarray}}
\newcommand{\Msol}{\ensuremath{M_{\odot}}}
\usepackage{color}

\begin{document}

\title{Cosmic String Loops as the Seeds of Super-Massive Black Holes}

\author{Sebastian F. Bramberger, Robert H. Brandenberger, Paul Jreidini, and Jerome Quintin}
\email{sebastian.bramberger@mail.mcgill.ca, rhb@physics.mcgill.ca, paul.jreidini@mail.mcgill.ca,
jquintin@physics.mcgill.ca}
\affiliation{Department of Physics, McGill University, Montr\'eal, QC, H3A 2T8, Canada}

\begin{abstract}
Recent discoveries of super-massive black holes at high redshifts indicate a possible tension
with the standard $\Lambda$CDM paradigm of early universe cosmology which has
difficulties in explaining the origin of the required nonlinear compact seeds which trigger 
the formation of these super-massive black holes. Here we show that cosmic string loops
which result from a scaling solution of strings formed during a phase transition in the
very early universe lead to an additional source of compact seeds. 
The number density of string-induced seeds dominates at high redshifts
and can help trigger the formation of the observed super-massive black holes.
\end{abstract}

\maketitle

\section{Introduction}
Super-massive black holes (SMBH) are among the most mysterious objects
in the Universe. Black holes are called ``super-massive'' if their
mass exceeds $10^6\,\Msol$, where $\Msol$ denotes the solar mass.
It is now believed that each galaxy harbors at least one super-massive 
black hole. The black hole nature of the massive object at the center
of our Milky Way galaxy has now been established without much
doubt by the precision observations of stellar orbits about it
(see e.g.\ \cite{Ghez}). The ultra-luminous quasars and active galactic
nuclei observed in other galaxies are believed to harbor black holes
(see e.g.\ \cite{Marta, Alexander} for recent reviews).

The origin of super-massive black holes is still somewhat of a
mystery. It is believed (see\ \cite{Marta}) that they result from
accretion of gas about massive seed objects. Three candidate
seed types are Population III stars with mass in the range
$10^2\,\Msol - 10^3\,\Msol$, dense matter clouds with mass
between $10^3\,\Msol$ and $10^6\,\Msol$, or compact objects
of mass between $10^2\,\Msol$ and $10^4\,\Msol$ formed by
the collision of old stellar clusters.
 
However, we must now explain the origin of the purported
seeds of the super-massive black holes. The  
recent observations of SMBHs of larger
masses and higher redshifts are leading to an increasing
tension with the standard paradigm of early universe cosmology
according to which the spectrum of primordial cosmological
fluctuations is approximately Gaussian with an almost scale-invariant
spectrum with a small red tilt. In the context of this model
nonlinearities form only at late times and there is not enough time to 
produce the nonlinear massive seeds which are required to seed 
SMBHs of mass greater or equal to $10^{9} \Msol$ at redshifts
of $6$ or higher (of which roughly 40 have been
discovered\ \cite{observs}). In particular, the recently discovered
black hole with mass $1.2 \times 10^{10}\,\Msol$ at redshift $z = 6.30$\ \cite{Nature}
is hard to explain in the context of the standard paradigm. 

Here we discuss the possibility that the compact seeds
which are required to be present at high redshifts are provided
by cosmic string loops.  String loops
are nonlinear seed masses which are present at arbitrarily
early times and which by gravitational accretion can seed
the objects which develop into SMBHs.

In the following we first give a very brief review
of the connection between cosmic strings and early
universe cosmology. We then compare the number densities
of nonlinear seeds in the vanilla $\Lambda$CDM 
cosmology with what is obtained when allowing for
the presence of a scaling distribution of cosmic strings.
We find that whereas the probability of finding
a seed in the range of Population III stars in the
vanilla $\Lambda$CDM model is too low to explain
the presence of SMBHs of the mass and redshift recently
discovered\ \cite{observs, Nature}, the presence of cosmic
strings easily solves this problem as long as the
mass per unit length $\mu$ of the strings obeys
the inequality $G \mu > 10^{-14}$.
Here, $G$ is Newton's gravitational constant and where we are
using natural units with $c=1$. We recover the appropriate
powers of $c$ when computing dimensionful observable quantites.

\section{Cosmic String Review}

Cosmic strings are linear topological
defects which are predicted in a large class of particle physics
models beyond the Standard Model (see\ \cite{CSrevs}
for reviews on cosmic strings). In particular,
cosmic strings are predicted to form after inflation
in many inflationary models, both models formulated
in the context of superstring theory\ \cite{Tye} and
in models based on supergravity\ \cite{Rachel}.
As first realized by Kibble\ \cite{Kibble},
causality arguments tell us that if Nature is described by a
particle physics model which admits cosmic string solutions,
then a network of cosmic strings will inevitably form in the
early universe and persist to the present time. The
distribution of cosmic strings consists of a network of
infinite strings and a set of string loops. Analytical
arguments tell us that the distribution of cosmic strings
will approach a ``scaling distribution'' in which $\xi(t)$,
the mean curvature radius and separation of the long string
network, will be of the order of the Hubble radius $t$. The
non-trivial dynamics of the long string segments will lead
to continuous loop production, and the distribution of
loops will also take on a scaling solution in which the
number density $n(R, t)$ of loops per unit radius $R$ 
is independent of time when $R$ is scaled to the Hubble radius $t$.

We will assume a simplified version of the cosmic string loop
scaling distribution according to which all loops formed at
time $t$ have the same radius $R_f(t)$,
\be
R_f(t) \, = \, \frac{\alpha}{\beta} t \, ,
\ee
where $\beta$ is the mean ratio of circumference to radius of a loop
(we will use the value $\beta = 10$), and
$\alpha$ is a constant whose value we shall take to be\ \cite{CSsimuls, Olum}
$\alpha \sim 0.1$. Loops are formed continuously in time. 
Note that whereas the value of $\beta$ has little uncertainty,
the value of $\alpha$ depends on details of the implementation
of cosmic string evolution simulations, and there is a large
range of possible values. We have taken a representative
value from the most recent cosmic string evolution simulations.

After formation, the number density of loops redshifts
due to the cosmic expansion. Loops also slowly decay by
emitting gravitational radiation\ \cite{gravrad}, and this
gives an effective lower cutoff for the range of string
loop radii at any given time. However, the string loops 
whose mean separation is comparable in comoving coordinates
to the separation $d_{\mathrm{gal}}$ of galaxies have a radius
larger than the gravitational radiation cutoff. They
are also formed before the time $t_{\mathrm{eq}}$ of equal matter
and radiation (the reader is invited the check these
statements). For such loops the number density per
unit radius is given by
\be \label{distrib}
n(R, t) \, = \,   N \alpha^{5/2} \beta^{-5/2} t_{\mathrm{eq}}^{1/2} t^{-2} R^{-5/2} \, ,  
\ee
where the number $N$ is determined by the number of
long string segments per Hubble volume.
Inserting for $t$ the present time $t_0$ gives the
comoving number density of these loops.

The trapped energy in cosmic strings leads to unique signals in cosmology.
Long string segments produce a conical discontinuity in space
which leads to lensing signals in CMB temperature
maps\ \cite{KS}, and to planar overdensities (called ``wakes'')
\cite{wakes} in the plane behind the moving strings which in
turn lead to direct B-mode polarization signals\ \cite{Holder1} and
to wedges of extra absorption in 21cm redshift maps\ \cite{Holder2} 
(see\ \cite{RHBCSreview} for an overview of these effects). 

String loops, on the other hand, accrete matter in a similar way
as a point mass (as long as distances large compared to the
loop radius are considered). It was at one point\ \cite{early}
postulated that cosmic string loops might be the seeds of
galaxies and galaxy clusters, without the need for Gaussian
fluctuations such as provided by inflation or its alternatives.
However, cosmic strings forming during a phase transition in
the early universe produce isocurvature fluctuations and hence
do not lead to coherent curvature perturbations on super-Hubble
scales, and hence do not generate acoustic oscillations in
the angular power spectrum of cosmic microwave anisotropies.
The discovery of these oscillations\ \cite{Boomerang} demonstrated
that the main source of fluctuations cannot be due to
cosmic strings. The most reliable limit on the cosmic string
tension in fact comes from detailed analyses of the CMB
angular power spectrum and yields\ \cite{SPT} 
(see also\ \cite{older})
\be
G \mu \, < \, 1.5 \times 10^{-7} \, .
\ee

String loops, however, may still play an important role
in cosmology. In a recent paper\ \cite{Globular} we
postulated that string loops may seed globular clusters.
Here we will study their role as possible seeds for
SMBHs.

In the following we first review the mechanism by
which compact seeds can lead to SMBHs. Then we
compute the expected distribution of seeds as a function
of mass and redshift in the $\Lambda$CDM model and
show that the number of seeds of mass required to
explain the most massive high redshift SMBHs is too
low. In the third subsection we then compute the
number density of seeds induced by cosmic string loops
and show that the string models can easily make up the
deficit of seeds for reasonable values of the tension $\mu$.

\section{Analysis}\label{Sec:analysis}

\subsection{Eddington accretion}

A compact seed evolves into a SMBH
by accreting gas at a rate that is
proportional to its bolometric luminosity.
In this way, an initial seed mass $M_i$ grows
into a final mass $M_f$ according to the equation
\be
M_f \, = \, M_i {\rm{exp}} \left( \frac{1 - \epsilon}{\epsilon} \frac{\Delta t}{t_*} \lambda \right)
\, ,
\ee
where $\Delta t$ is the time interval during which accretion
takes place, $t_* = 4.5 \times 10^8~{\rm yrs}$ is the Eddington
time, $\lambda$ is the ratio of the bolometric luminosity
to the Eddington luminosity\ \cite{Eddington},
and $\epsilon$ is the efficiency of radiative emission which
typically depends on the black-hole spin, and
we shall take $\epsilon = 0.1$ (see\ \cite{Efficiency}).
Expressing time in terms of redshift this equation can be
re-written as
\bea
M(z_i)  &=&  M(z) \\
&& \times {\rm{exp}} \left[ -  \frac{1 - \epsilon}{\epsilon}
\frac{t_0}{t_*} \left(\frac{1}{(1 + z)^{3/2}} - \frac{1}{(1 + z_i)^{3/2}} \right) \lambda \right]
\, , \nonumber
\eea
where $M(z_i)$ is the initial seed mass at redshift $z_i$ and $M(z)$
is the mass at the final redshift $z$.

\begin{figure*}
\includegraphics[scale=0.75]{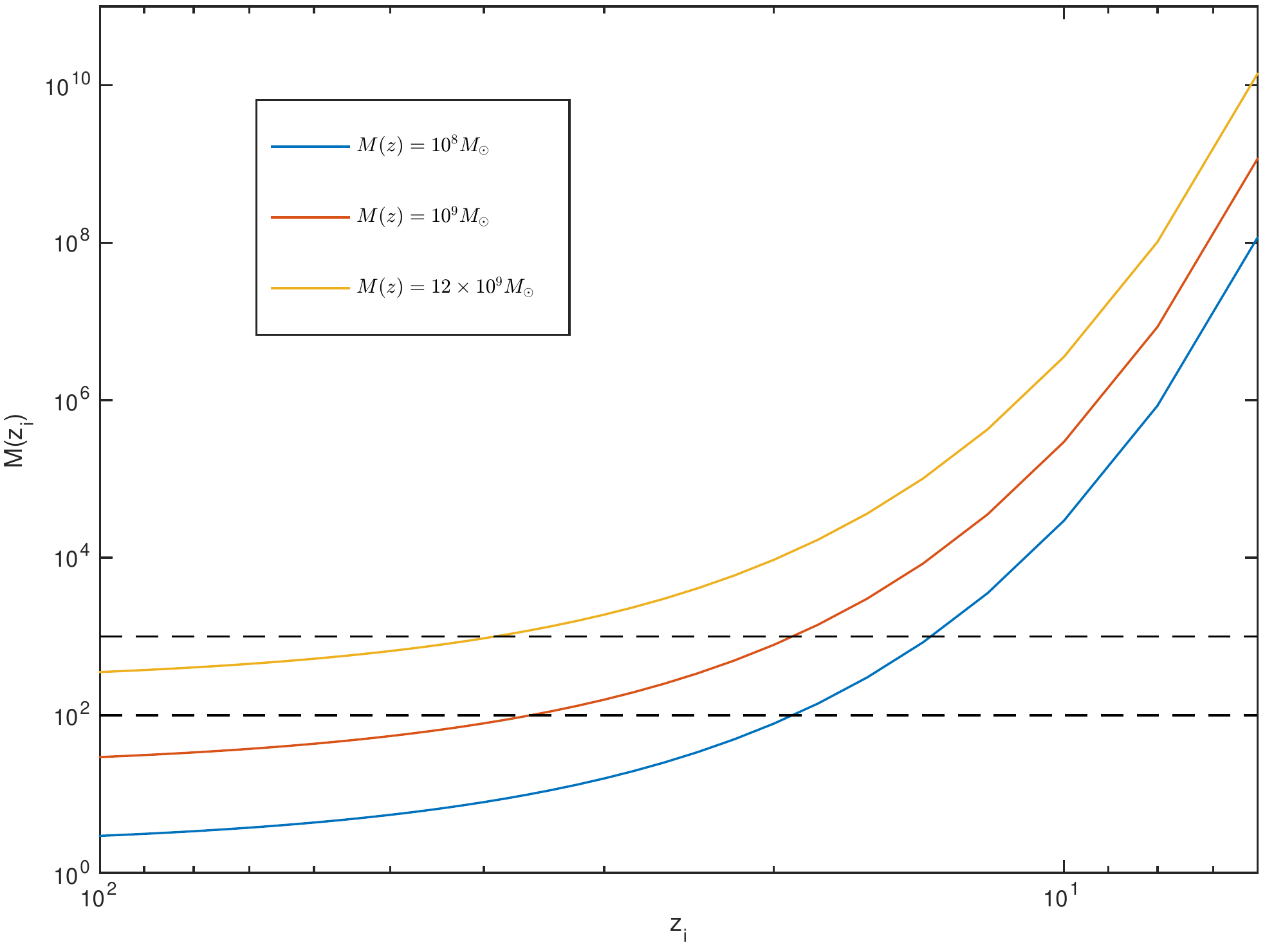}
\caption{Seed mass required to produce a SMBH at redshift $6.3$ of
masses $M=1.2 \times 10^{10}\,\Msol$, $M=10^9\,\Msol$, and $M = 10^8\,\Msol$ as a 
function of the redshift $z_i$ when the compact seeds is
produced, assuming $\lambda = 1$. The horizontal axis is the redshift $z_i$, and the vertical
axis gives the mass (in solar mass units). The two dashed
black lines show the mass range of the postulated
Population III star seeds.} \label{Fig:Eddington}
\end{figure*}

Fig.\ \ref{Fig:Eddington} shows the seed mass $M(z_i)$ required to form a SMBH at
final redshift $z = 6.3$ (the redshift of the recently discovered object)
and for final SMBH masses of $M(z) = 1.2 \times 10^{10}\,\Msol$ (the
estimated mass of the discovered object), $M(z) = 10^9\,\Msol$ and
$M(z) = 10^8\,\Msol$, assuming $\lambda = 1$. The horizontal axis is $z_i$ and the
vertical axis is the mass. The two dashed horizontal lines at $M = 10^2\,\Msol$
and $M = 10^3\,\Msol$ give the mass range of the postulated
Population III star seeds. The graph shows that seeds in this mass
range had to have been present at redshift of greater than 40 in
order to grow into the recently observed SMBH with mass of
$1.2 \times 10^{10}\,\Msol$. The seeds for $10^9\,\Msol$ objects
needed to have been present in the required number density  
by redshift of $z = 20$. As we will show in the next subsection,
we do not expect the $\Lambda$CDM model to yield any nonlinear 
objects of the required masses at these high redshifts.
The initial redshift at which the larger seed masses postulated for the two other conventional
sources of SMBH formation needed to have been present is smaller, but
since the mass is larger, this does not necessarily make them easier to
produce in the standard paradigm of early universe cosmology, as we
will see in the following subsection.

The assumption that $\lambda=1$ implies that SMBHs accrete mass
at the Eddington rate. If the accretion process were sub-Eddington,
it would force the seeds to have been present at even higher redshifts
and thus increase the tension with $\Lambda$CDM. For simplicity, we consider
Eddington accretion as a limiting case, keeping in mind that more
massive seeds may be required at even higher redshifts. Obviously,
if the accretion were super-Eddington, there the tension with
the standard cosmological model would be less (see\ \cite{Volonteri:2014lja}).
For studies trying to constrain $\lambda$, see e.g.\ \cite{Volonteri_more}.

\subsection{Density of Seeds in the $\Lambda$CDM Model}

\begin{figure*}
\includegraphics[scale=0.75]{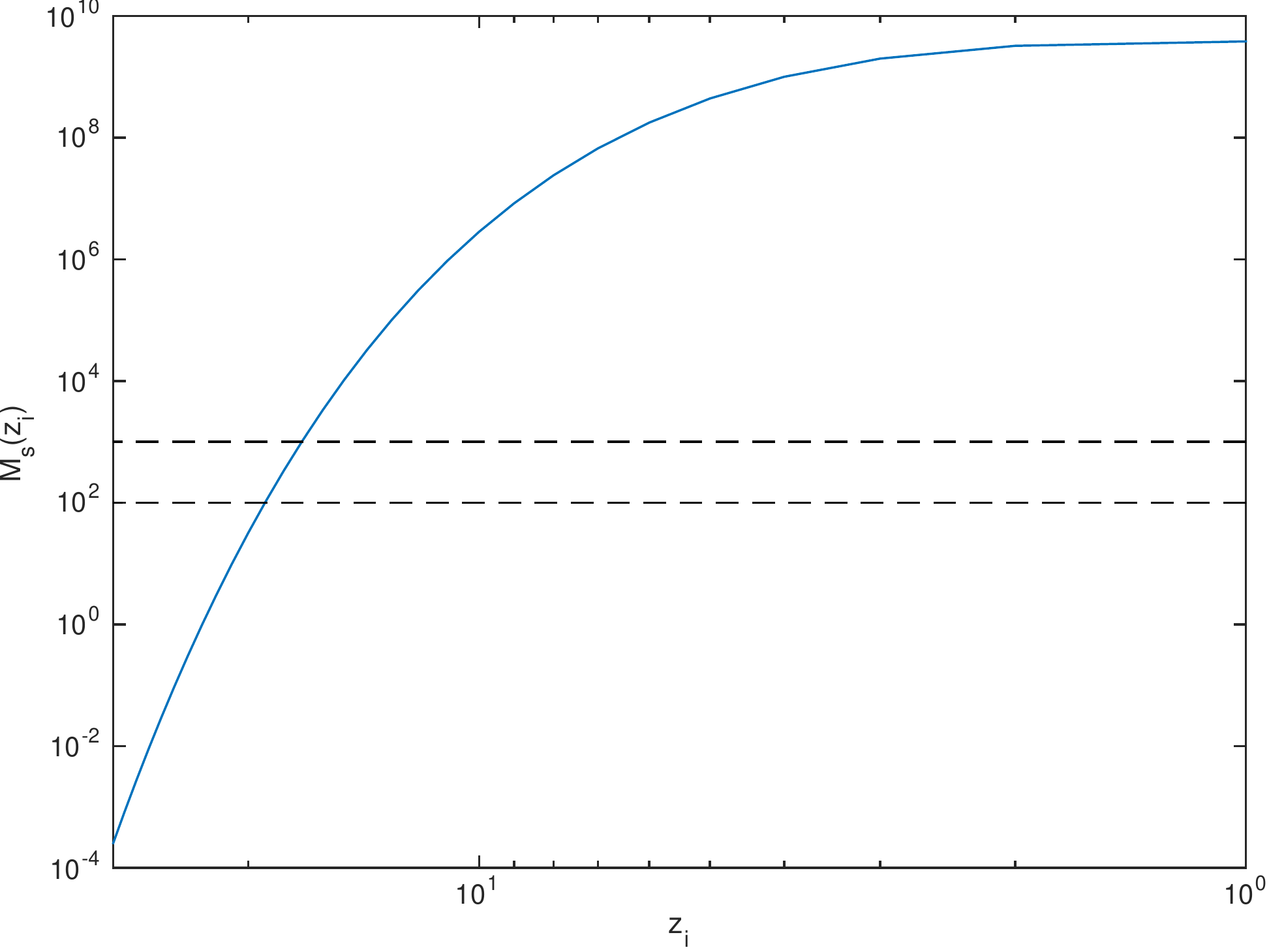}
\caption{Mass $M_s(z_i)$ of seeds from Gaussian
fluctuations as a function of redshift $z_i$ (horizontal axis).
The vertical axis is the mass (again in solar mass units). As in
the previous figure, the two horizontal dashed black lines indicate
the mass range of Population III stars.}\label{Fig:gauss_fluct}
\end{figure*}

For Gaussian fluctuations, it is exponentially unlikely to
obtain a nonlinear fluctuation of mass $m$ if the root mean
square value of the dimensionless density power spectrum
$\sigma(m)$ is smaller than $1$. More precisely, the mass 
function $n(m, z)$ at redshift $z$ is given by (see e.g.\ \cite{Sheth})
\be \label{massfct}
n(m, z) m \, = \, \rho m^{-1} \left[ \frac{d \log m}{d \log \nu} \right]^{-1} \nu f(\nu) \, ,
\ee
where $\rho$ is the background density and
$\nu$ is the excess over the r.m.s. value required to
form a nonlinear object, i.e.
\be \label{nu}
\nu \, \equiv \,  \left( \frac{\delta_c(z)}{\sigma(m)} \right)^2 \, .
\ee
The number $\delta_c(z)$ and its redshift dependence depend
slightly on the background cosmology. For our analysis
we will take the value $\delta_c = 1.7$ (independent of
redshift)\ \cite{Sheth}. This corresponds to neglecting the
effects of dark energy. This is a good approximation for our
problem where we are interested in high redshifts where
dark energy does not have an important effect. For the
function $f(\nu)$ we will use the Press-Schechter\ \cite{PS} form
\be \label{eff}
\nu f(\nu) \, = \left( \frac{\nu}{2} \right)^{1/2} \frac{1}{\sqrt{\pi}} e^{- \nu / 2} \, .
\ee

We are interested in mass ranges for which the corresponding
wavelength is smaller than the Hubble radius at $t_{\mathrm{eq}}$ and
which are hence in the region where the power spectrum
increases only logarithmically as $m^{-1}$. Specifically,
we make the ansatz
\be \label{sigma}
\sigma^2(m) \, = \, A \log(m_c / m) (1 + z)^{-2} \, ,
\ee
where the last factor is the linear perturbation theory growth
of the amplitude of the power spectrum. If we choose $m_c$
to be the mass scale where $\sigma(m) = 1$ today then we
can set $A = 1$.

Inserting\ \eqref{sigma},\ \eqref{eff}, and\ \eqref{nu} into the expression
\eqref{massfct} for the mass function, we can determine the mass $M_s(z)$
of seeds which have the comoving separation of galaxies at redshift $z$.
This amounts to solving the equation
\be
\label{eq:dn1}
d_{\mathrm{gal}}^3 n(m, z) m \, = \, 1 \, 
\ee
for the mass $m$, where $d_{\mathrm{gal}}$ is the comoving separation of galaxies.
We take $d_{\mathrm{gal}}=1~\mathrm{Mpc}$ independent of redshift, but a more 
sophisticated analysis could introduce some redshift dependence, although this 
may no be well-constrained at high redshifts.
We find the following transcendental equation from Eq.\ \eqref{eq:dn1},
\bea
1 \, &=& \, d_{\mathrm{gal}}^3 \rho \delta_c \left( \frac{1}{2 \pi} \right)^{1/2} (z + 1) \\
& & \times \frac{1}{\log^{3/2} (m_c / m)} m^{-1}
{\rm exp}\left[- \frac{(z + 1)^2 \delta_c^2}{2 \log(m_c / m)}\right] \, . \nonumber
\eea

Solving for $m$, the resulting function $M_s(z)$ is plotted in Fig.\ \ref{Fig:gauss_fluct}.
Note the exponential decrease of the seed mass at high redshifts. This implies that the Gaussian
fluctuations in the standard $\Lambda$CDM model have trouble explaining
the origins of the massive compact seeds which are required to explain
the formation of the highest redshift and most massive observed SMBHs.
At redshifts of 20 and higher, the mass of the predicted seeds with the number
density of galaxies is smaller than $10^2\,\Msol$.
In the following subsection we find that cosmic string loops can come to
the rescue.

\subsection{Density of Seeds induced by Cosmic String Loops}

Here we compute the number density of string-induced
compact seeds. In the range of loop radius $R$ which we
are interested in the loops are already present at
$t_{\mathrm{eq}}$. Taking the linear perturbation theory growth
in mass, the mass of the nonlinear seed produced by
a loop of radius $R$ at redshift $z$ is
\be
M(R, z) \, = \, \frac{z_{\mathrm{eq}} + 1}{z + 1} \beta \mu R \, ,
\ee
where $z_{\mathrm{eq}}$ is the redshift at $t_{\mathrm{eq}}$. Taking
into account the Jacobian of the transformation between
$R$ and $M$, the resulting seed mass function is
\bea
n(M, z) M \, &=& \, N \beta^{-1} \alpha^{5/2} \left(\frac{z_{\mathrm{eq}} + 1}{z + 1}\right)^{3/2} \\
& & \times\, t_{\mathrm{eq}}^{1/2} t_0^{-2} \mu^{3/2} M^{-3/2} \, , \nonumber
\eea
where $n(M, z) M$ gives the number density in comoving
coordinates of seeds with mass in the range $[M, 2M]$.

\begin{figure}
\includegraphics[scale=0.43]{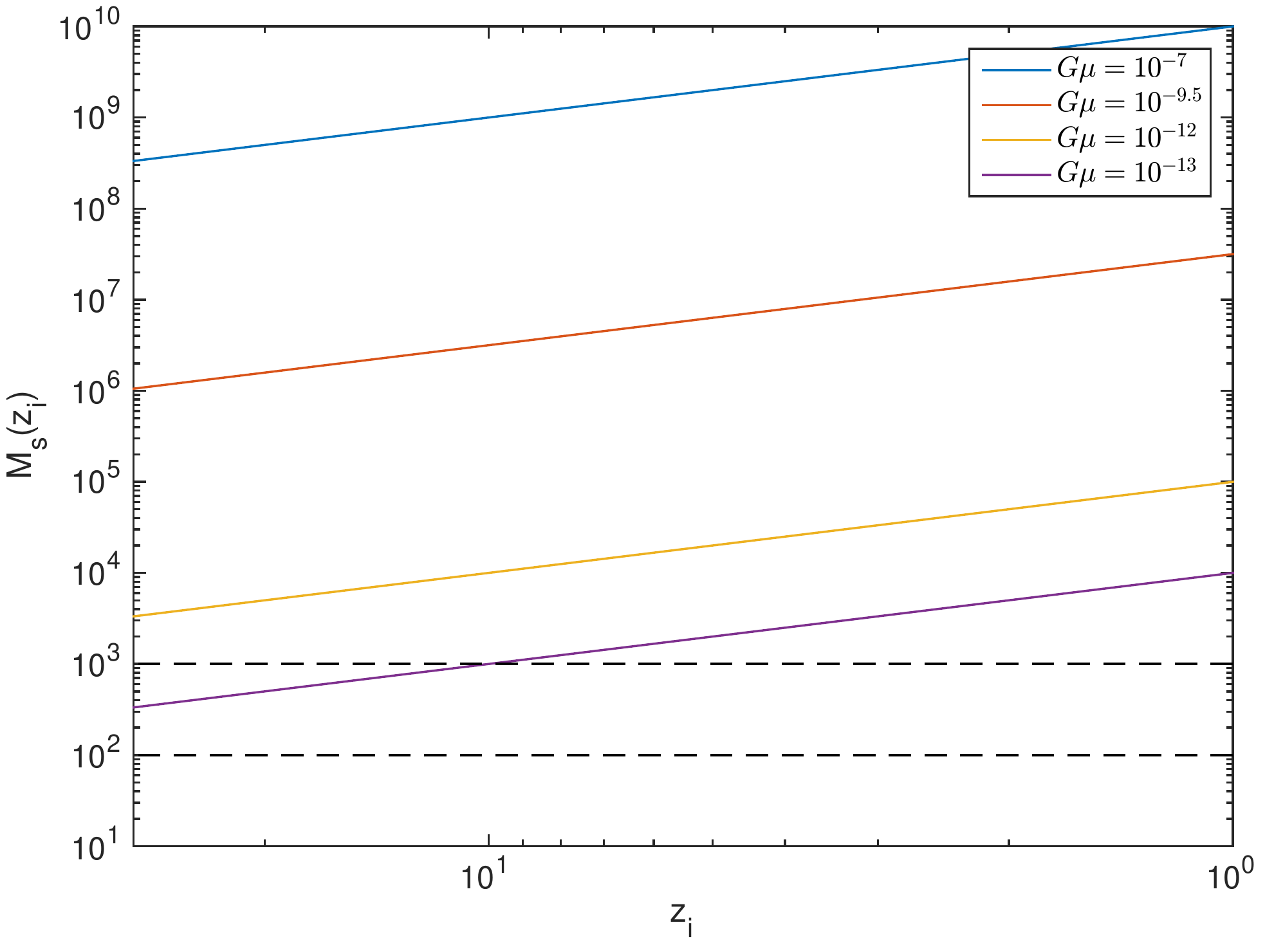}
\caption{Mass $M_s(z_i)$ of seeds induced by cosmic string
loops which at redshift $z_i$ have the mean comoving
separation of galaxies. The curves are for values
$G \mu = 10^{-7}$ (top curve) and $G \mu = 10^{-9.5}$ (bottom curve).
The first value is just below the current upper bound, and the
second value is the one which gives a good fit to the
mass function of globular clusters. The masses on the vertical axis
are once again in $\Msol$.} \label{globfig}
\end{figure}

We can now compute the mass $M_s(z)$ at redshift $z$ of seeds
with a mean separation of $d_{\mathrm{gal}}$. This is determined
by
\be
n(M_s, z) M_s d_{\mathrm{gal}}^3 \, = \, 1 \, ,
\ee
and yields
\be
M_s(z) \, = \, N^{2/3} \beta^{-2/3} z_{\mathrm{eq}}^{-1/2} G \mu \frac{t_0}{G}
\left( \frac{d_{\mathrm{gal}}}{t_0} \right)^2 \frac{z_{\mathrm{eq}} + 1}{z + 1} \, .
\ee
The only dimensionful number which enters the above
expression is $t_0/G \simeq 10^{23}\,\Msol$.

\begin{figure*}
\includegraphics[scale=0.75]{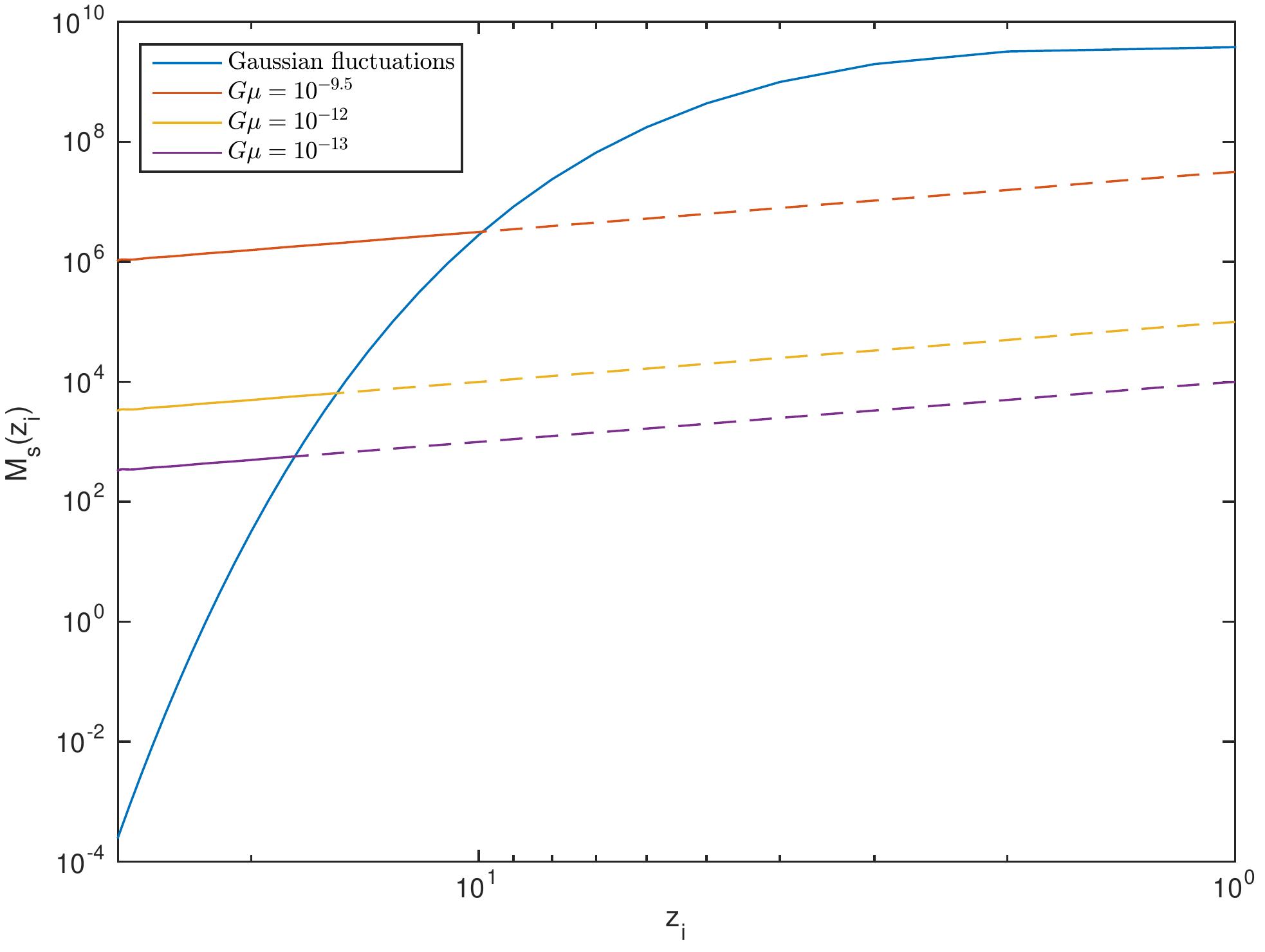}
\caption{Mass $M_s(z_i)$ of seeds from both Gaussian
fluctuations and from cosmic string loops. It is evident that
the cosmic strings dominate the seed distribution at high
redshifts. The units of mass on the vertical axis are in $\Msol$.} \label{Fig:combined}
\end{figure*}

In Fig.\ \ref{globfig} we plot the mass $M_s(z_i)$ (vertical axis) of
seeds which have the mean comoving separation of galaxies
as a function of the initial redshift $z_i$ (horizontal axis)
for four values of the dimensionless string tension $G \mu$.
Comparing with the curves in Fig.\ \ref{Fig:Eddington}, we see that
for the values $G \mu = 10^{-7}$ (which is just below the current
upper bound on $G \mu$ from CMB data), and $G \mu = 10^{-9.5}$
(for which string loops can explain the origin of globular clusters
\cite{Globular}) the seed masses present in the correct number density exceed 
at redshifts of $20$ or higher the mass of Population III stars, and hence
for these values of $G \mu$ there are more than enough seeds
to develop into SMBHs. Demanding that string loops provide
the right number density of seeds of mass $10^3\,\Msol$ and
$10^2\,\Msol$ leads to values of $G \mu$ of
$10^{-12}$ and $10^{-13}$, respectively. The seed masses
for these values of $G \mu$ are also shown in Fig.\ \ref{globfig}.

Comparing with the results of Fig.\ \ref{Fig:Eddington}, we see that a seed
mass of about $10^3\,\Msol$ is required to be present at redshift 40
to explain the recently discovered SMBH. 
We see that for values of $G \mu > 10^{-12}$ there will be
a sufficient number of such seeds present. The less stringent
requirements of having $10^3 \Msol$ or $10^2 \Msol$ seeds present
at redshift of 20 (which will explain $10^9 \Msol$ and $10^8 \Msol$
SMBH masses at redshifts of about 6 are satisfied if
$G \mu > 2 \times 10^{-13}$ and $G \mu > 2 \times 10^{14}$, respectively.

Fig.\ \ref{Fig:combined} is an overlay of the results of
Figs.\ \ref{Fig:gauss_fluct} and\ \ref{globfig}, namely
the mass of seeds separated by a mean galactic distance, taking
into account both seeds formed from Gaussian fluctuations and
from string loops. As expected, the string loops dominate at
high redshift. The crossover redshift depends on $G \mu$.
For the entire range of values of $G \mu$ which are of interest
to us ( $10^{-13} <  G \mu < 10^{-7}$) the string loops dominate
the seed distribution at redshifts of 20 and higher.

\section{Conclusions}

The discovery of super-massive black holes at redshifts
greater than $z = 6$ and with masses greater than
$10^9 \Msol$ is challenging to explain in the context of
a pure $\Lambda$CDM cosmology since the number
density of the required seeds is not predicted to be
high enough (unless super-Eddington accretion is invoked). 
In this note we have suggested that cosmic string 
loops might be the seeds about which super-massive black holes form. Cosmic
string loops lead to massive compact objects already at high redshift,
and can thus provide the massive seeds at high redshifts which
are missing in the vanilla $\Lambda$CDM model. We
have computed the masses of seeds with the correct
number density to explain the origin of one SMBH per galaxy
both with and without cosmic strings.
We note that the one SMBH per galaxy condition is more
stringent then necessary since not all galaxies need to host a
SMBH at high $z$ to host one today\ \cite{Menou:2001hb}.
Still, we find that even cosmic strings with a tension significantly lower
than the current upper bound can provide enough
seeds at early times.

We are thus postulating that cosmic string loops can play
the role which Population III stars or dense early gas clouds
are thought to play in the current SMBH formation scenarios
\cite{Marta}. Cosmic string loops lead to compact nonlinear
structures with falling rotation curves. This makes string 
loops appealing seed candidates for the formation of SMBH. A possible
problem, however, is that the gravitational potential
may not be deep enough to allow the cooling and star formation\ \cite{Loeb}
required to obtain the compactness needed to get an actual
black hole.

In the absence of knowing the exact mechanism by which
the seed object leads to the formation of a SMBH,
and without more information about the statistics of SMBH, we cannot
give a precise value for the string tension $G \mu$ for which
our mechanism works best. At the end of Sec.\ \ref{Sec:analysis} we have given
some representative values. They all lie comfortably below
the upper bound from current observations. 

We do not claim that our cosmic string loop model is the only
way to supplement the standard cosmological paradigm in order
to provide a way to obtain large mass high redshift SMBHs.
Another example is the recent suggestion\ \cite{Spergel} that a
small fraction of the dark matter is ultra-strongly interacting and
can undergo gravothermal collapse at early times, leading to
compact seeds.

\section*{Acknowledgments}
We wish to thank Julie Hlavacek-Larrondo, Gil Holder, and
Marta Volonteri for valuable discussions and encouragement.
RB is supported by an NSERC Discovery Grant, and by funds 
from the Canada Research Chair program.
JQ acknowledges the Fonds de recherche du Qu\'{e}bec - Nature et
technologies (FRQNT) for financial support.

\end{document}